\documentclass[fleqn,10pt]{JLA_article} 

\usepackage[english]{babel} 

\usepackage{lipsum} 
\usepackage{lineno}
\usepackage{soul} 
\newcommand{\name}{RiPPLE\xspace}
\usepackage{xspace}
\usepackage{graphicx}   
\usepackage{amssymb}
\usepackage{subcaption}

\newcommand{\alss}{ALSs\xspace}
\usepackage{multirow}

\usepackage{hyperref} 
\hypersetup{hidelinks,colorlinks,breaklinks=true,urlcolor=color2,citecolor=color1,linkcolor=color1,bookmarksopen=false,pdftitle={Title},pdfauthor={Author}}

\PaperTitle{\name: A Crowdsourced Adaptive Platform for Recommendation of Learning Activities}

\Authors{Hassan Khosravi\textsuperscript{1}, Kirsty Kitto\textsuperscript{2}, Joseph Jay Williams\textsuperscript{3}} 
\affiliation{Correponding author \textsuperscript{1}\textit{
Email: h.khosravi@uq.edu.au
Address: Institute for Teaching and Learning Innovation, The University of Queensland, St Lucia QLD 4072, Australia.
ORCID: 0000-0001-8664-6117
 }} 
\affiliation{\textsuperscript{2}\textit{
Address:  Connected Intelligence Centre, University of Technology Sydney, 15 Broadway, Ultimo NSW 200, Australia.
ORCID: 0000-0001-7642-7121}}

\affiliation{\textsuperscript{3}\textit{
Address: Department of Computer Science, University of Toronto, 40 St. George Street, Room 7224,
Toronto, ON M5S 2E4, Canada
 }} 


\Submitted{28/11/18}
\Accepted{03/09/2019}
\Published{}

\Volume{6}
\Number{3}
\Pages{1---10}
\Doi{xxx-xxx-xxx}
\Notesname{Notes for Practice} 

\note{Adaptive learning systems (ALSs) dynamically adjust the level or type of instruction based on individual student abilities or preferences
to provide a customized learning experience.}
\note{However, ALSs require access to a large repository of learning resources, which are commonly created by domain experts. This makes them expensive to develop and challenging to scale across different domains.}
\note{We introduce an ALS called RiPPLE that uses a crowdsourcing approach to engage students in creating, moderating, and evaluating learning resources. This both reduces the cost of content generation and has the potential to foster higher-order learning across many domains.}
\note{Initial results suggest that using RiPPLE may lead to measurable learning gains and that students perceived the platform as beneficially supporting their learning.}

\Abstract{
This paper presents a platform called \name (Recommendation in Personalised Peer-Learning Environments) 
that recommends personalized learning activities to students based on their knowledge state from a pool of crowdsourced learning activities that are generated by educators and the students themselves. \name integrates insights from crowdsourcing, learning sciences, and adaptive learning, aiming to narrow the gap between these large bodies of research while providing a practical platform-based implementation that instructors can easily use in their courses. This paper provides a design overview of \name, which can be employed as a standalone tool or embedded into any learning management system (LMS) or online platform that supports the Learning Tools Interoperability (LTI) standard. The platform has been evaluated based on a pilot in an introductory course with 453 students at The University of Queensland. Initial results suggest that the use of the \name platform led to measurable learning gains and that students perceived the platform as beneficially supporting their learning.}

\begin{document}

\flushbottom 

\maketitle 


\thispagestyle{fancy} 


\section{Introduction}
Educators continue to face significant challenges in providing high-quality postsecondary instruction in large online or on-campus classes \cite{mulryan2010teaching}. A significant portion of these challenges emerge from high levels of diversity in learners' academic ability \cite{banks2005teaching}. Teach-to-the-middle instruction is most commonly used; however, this method does not meet the needs of learners who differ significantly from the norm.
Adaptive learning systems (\alss) \cite{park2003adaptive} provide a potential solution to this problem. \alss make use of data about students, learning processes, and learning products to dynamically adapt learning content and activities to suit students' individual abilities or preferences. A growing body of knowledge provides evidence about the effectiveness of \alss in supporting student learning \cite{anderson1995cognitive,vanlehn2011relative,ma2014intelligent}. 

The most successful adaptive platforms tend to focus on a specific domain (see Section~\ref{sec:relatedWork}) and also require a tremendous investment of time by experts during the development of curriculum content, which makes it difficult to scale these approaches across many domains. As a viable alternative, we see great promise in trying to leverage ideas from crowdsourcing and learning science to create adaptive systems by having students themselves generate content that can be adaptively served. The benefits of engaging students in content creation are twofold. First, students transform from passive recipients of content to active creators of course material. Previous studies have reported that placing the responsibility of content creation in the hands of students reinforces and deepens their understanding of the course content \cite{barak2004line, draper2009catalytic}, highlights the significance of representing their work in a clear and logical fashion \cite{Denny2008}, encourages reflection on the course objectives \cite{purchase2010quality}, and enhances their conceptual understanding \cite{bates2012}. Second, the creativity of  students themselves is used to develop large repositories of learning resources. Previous studies have demonstrated that students can create high-quality learning resources that meet rigorous judgmental and statistical criteria \cite{walsh2018formative,tackett2018crowdsourcing,denny2009students,galloway2015doing, bates2014assessing}.

This paper discusses the design and implementation of a platform called \name  (Recommendation in Personalised Peer Learning Environments)\footnote{\url{http://ripplelearning.org}}
that adopts a students-as-partners approach \cite{matthews2017five} to provide adaptive learning functionalities while enabling learners to be more cognitively involved in their learning. The platform provides easily accessible data, which will help the learning analytics (LA) community research a number of topics of interest, including the type of content generated by students for their peers, the way students interact with peer-generated questions, and student use of recommendation systems and open learner models to direct their learning.  

The platform was initially evaluated based on a pilot in an introductory course with 453 students at The University of Queensland. Initial results suggest that the use of the \name platform may lead to significant and measurable learning gains; the mid-semester score of the experimental group using the platform was approximately 8 percentage points 
(total effect size: $d= 0.54$ \cite{cohen1992power}, $p < .001$) higher than the control group that was not using the platform. Additionally, a student survey ($N=55$) shows that a large fraction of the respondees believe that \name helped them to study more efficiently and effectively. 

In what follows, we introduce \name to the LA community.  Section~\ref{sec:relatedWork} provides a background for the use of adaptive learning and crowdsourcing in education. Section~\ref{sec:platform} introduces the \name platform, and Section~\ref{sec:evaluation} evaluates the platform in an authentic learning context. Finally, Section~\ref{sec:conclusions} presents concluding remarks.
We start with a consideration of \alss and how they can help students navigate their way through large amounts of learning content.

\section{Background: Adaptive Learning and Crowdsourcing in Education} \label{sec:relatedWork}

\alss \cite{park2003adaptive} dynamically adjust the level or type of instruction based on individual student abilities or preferences to provide an efficient, effective, and customized learning experience. At a high level of generality, there are two main classes of \alss. The first class, commonly referred to as intelligent tutoring systems (ITSs) \cite{anderson1985intelligent}, uses AI-based algorithms to replicate the support that is often provided by a tutor by providing personalized step-by-step guidance in solving a problem. Carnegie Learning's MATHiaU \cite{ritter2015carnegie} is an established example of this class of \alss. The second class of \alss focuses on adaptively recommending learning activities from a large repository of learning resources to a student to match their current learning ability. Pearson's MyLabs (using Knewton \cite{Ferreira2016white} for its adaptive functionality) and McGraw-Hill's LearnSmart and ALEKS \cite{falmagne2006assessment} are established examples of this class of \alss. \name is also representative of this second class. 

A consistent and growing body of knowledge provides evidence about the effectiveness of both classes of \alss \cite{anderson1995cognitive,vanlehn2011relative,ma2014intelligent}. For example, a comprehensive meta-review by \citeA{vanlehn2011relative} reported that on average, students using ITSs have a learning gain effect size of $d= 0.76$  relative to classroom teaching without tutors. For \alss that focus on recommending learning resources, \citeA{yilmaz2017effects} and \citeA{ mojaradstudying} have reported improvement on student performance while using ALEKS or the popular ASSISTments Ecosystem \cite{heffernan2014assistments}. 

At a high level of generality, many \alss rely on the following interacting components \cite{essa2016possible}: 
\begin{enumerate}[noitemsep]
\item \textit{Domain model}: A knowledge space modelling what the students need to know. The domain model is commonly presented as a set of knowledge units that are ``elementary fragments of knowledge for the given domain" \cite{brusilovsky2012adaptive}.
\item \textit{Learner model}: An abstract representation of students, often ``overlaying" their knowledge state on the knowledge space defined in the domain model \cite{brusilovsky2007user}. The learner model may estimate a student's ability level on different knowledge units based on their performance and interactions with the system. Importantly, open learner models \cite{bull2010open}, which are learner models that are externalized and made accessible to students or other stakeholders, can be particularly effective in helping students learn \cite{bodily2018open}.

\item \textit{Content repository}: A repository of learning resources that may include assessment-based and learning-based items designed to help the learner acquire the knowledge represented by the domain model. Each learning resource is tagged with knowledge units defined in the domain model.
\item \textit{Recommender engine}: An engine that utilizes information from the learner model and the content repository to select learning activities for each student that will be most likely to advance their learning of the domain knowledge.
\end{enumerate}

Commonly, \alss are developed using the publisher model \cite{oxman2014white}. In this model, the platform is designed with pre-existing learning activities, often based on textbooks from a publisher. Pearson’s MyLabs (using Knewton \cite{Ferreira2016white} for its adaptive functionality) and McGraw-Hill’s LearnSmart and ALEKS \cite{falmagne2006assessment} are established examples of this model. The publisher model has been successful in K--12, where course content follows a simpler structure and often has to comply with national standards. However, higher education has been slow to embrace these systems, with adoption mostly restricted to research projects \cite{essa2016possible}. The focus on specific restricted domains, limited flexibility for educators to tailor the learning activities to their context, and the high costs associated with the use of these platforms have all contributed to their low adoption in higher education.

Responding to these limitations, an alternative has been established, referred to as the platform model \cite{oxman2014white}. The platform model provides a content-agnostic system infrastructure that enables educators to develop and author the content of their course. Smart Sparrow \cite{Sparrow2016} and many learning management systems (LMSs), such as Desire2Learn, Loudcloud, and edX, which incorporate adaptive functionality into their course-building tools, follow this model. The platform model is relatively new and mostly suffers from an operational limitation rather than a technological one; implementing adaptivity in a course requires a large number of new learning activities and object tagging, introducing significant overhead for teaching staff in both time and training. To overcome the limitations of both of these models, \name leverages ideas from crowdsourcing in education \cite{solemon2013review} by having students themselves generate and evaluate content that can then be adaptively served.

The use of crowdsourcing in education alongside insights from the students-as-partners approach \cite{matthews2017five} makes way for respectful, mutually beneficial learning partnerships where students and staff work together. Successful examples of such partnerships have led to co-creation of curricula \cite{bovill2013students}; marking criteria \cite{meer2014co}; and assessment items via the popular PeerWise platform \cite{Denny2008}, which has inspired \name. The use of crowdsourcing in \alss is also beginning to receive attention. For example,  \citeA{heffernan2016future} propose employing crowdsourcing within the popular ASSISTments platform; 
\citeA{williams2016axis} present an Adaptive eXplanation Improvement System (AXIS) that uses crowdsourcing to generate, revise, and evaluate explanations as learners solve problems; and \citeA{karataev2017adaptive} propose a framework that combines concepts of crowdsourcing, online social networks, and adaptive systems to provide personalized learning pathways for students. However, this preliminary work is yet to realize the full potential offered by crowdsourcing in \alss or more broadly in education. An important motivator in developing \name is to provide support for ethical and low-cost educational research on the use of crowdsourcing in education and \alss.

\section{The \name Platform}\label{sec:platform}

In this section, we provide an overview of the main functionalities of \name. Section~\ref{sec:overview} provides an overview of how a new \name offering can be created. Section~\ref{sec:creation} presents how learning activities are created, attempted, and evaluated in \name. Section~\ref{sec:learnerModel} then introduces the open learner model used by \name, which enables students to view an abstract representation of their knowledge state. Section~\ref{sec:recommendation} demonstrates the available features for resource selection and recommendation, and Section~\ref{sec:studentProfiles} describes the information available under personal profiles and course reports. 
Prototypes of the platform, demonstrating the student view and the instructor view, and further information are available on the web page of the platform\footnote{\url{http://ripplelearning.org}}.

\subsection{Creation of a \name Offering}\label{sec:overview}
\name can be integrated into many popular LMSs (including Blackboard, Moodle, and Canvas) using the Learning Tools Interoperability (LTI) standard. Once integrated into an LMS, \name can be added as an LTI tool (via a link) to any course within that system. \name supports two types of roles for users: instructors and students. The role of a user is determined using the LTI standard based on their role in the LMS course that is hosting \name. For example, assuming that \name is hosted in a course on Blackboard, users with the Instructor, Teaching Assistant, or Grader role in that course would be given the instructor role on \name. Similarly, users with the Student, Guest, or Observer role in that course on Blackboard would be given the student role on \name. Once a tool link has been added to an LMS course, a user with the instructor role can click on the link to create a \name offering for the course. 

 The university name, course code, course name, course semester, and teaching start date are all automatically completed based on the information of the course captured in the LMS. The instructor is required to enter a set of topics, representing the knowledge units, which define the knowledge space of a course. They can do this by importing topics from other \name offerings, creating new ones, or modifying an existing list of topics (e.g., renaming, deleting, or changing the order of the knowledge units). The list of topics may be altered throughout the semester. 

\name can also be used as a standalone system without integration into an LMS. In this case, the instructor would (1) create a personal account in order to create a \name offering; (2) manually enter the university name, course code, course name, course semester, teaching start date, and course topics; and (3) add other users (e.g., students and instructors) to the created \name offering. Two methods are supported for adding other users to an offering: by invitation, where users receive an email to join the offering, or by self-enrolment, where users receive an access code that can be used to manually sign up for the course. 

Once a \name offering has been created, an instructor can import resources from other \name offerings. This enables instructors to import resources from their past offerings as well as share resources with other instructors who are teaching similar courses. The resource import page allows instructors to search for resources from other offerings based on many options, including university name, course name, offering ID, topics, resource rating, resource type, and keywords.

\subsection{Content Creation and Evaluation}\label{sec:creation}
\name integrates insight from classical and contemporary models of learning to engage students in activities across many of the higher-level objectives of the cognitive domain of Bloom's taxonomy \cite{bloom1956taxonomy}. For example, \name enables students to author learning activities that help them think carefully about the concepts and learning outcomes of the course, formulate distractors for multiple-choice questions (MCQs) (which requires them to analyze misconceptions their peers might have), and explain their understanding of a concept to other students who chose the wrong answer.  More recent models of learning also provide evidence that engaging students in elaborating on the learning content \cite{pressley1992encouraging,king1992facilitating}, such as developing evaluative judgment \cite{tai2018developing}, can increase learning. In \name, these elaborations can take many forms, including clarifying an idea, constructing an original explanation of a concept, judging the quality of an activity, or comparing and contrasting an activity with other activities. 

\name enables students and instructors to create, attempt, and evaluate a wide range of learning resources that are tagged with one or more topics pre-assigned by users who have instructor status in the course. The resources that can be created currently include MCQs, worked examples, and general notes. Creating an MCQ includes (1) developing the body of the question, (2) tagging it with one or more topics, (3) developing the multiple-choice answers, (4) nominating the correct answer, and (5) writing an explanation of the solution. Creating a worked example includes (1) developing the body of the question, (2) tagging it with one or more topics, (3) developing a series of steps that work through answering the question, and (4) developing a final solution for the question. Creating a general note is more flexible and can be used to develop a broad range of learning resources, including mini-lessons (teaching a concept or applying a theoretical concept to real-world examples), summary notes, and cheat sheets. In all cases, text, tables, images, videos, and scientific formulas can be used to create learning resources. Authors can also view their learning resources before submitting and edit their submitted resources at a later time using \name. 

An important assumption made by \name is that students, as non-experts, can create high-quality resources. While we were not able to find longitudinal or meta-reviews that support this claim, there seems to be adequate evidence suggesting that students can create high-quality learning resources that meet rigorous judgmental and statistical criteria \cite{walsh2018formative,tackett2018crowdsourcing,denny2009students,galloway2015doing, bates2014assessing}. In fact, students as authors of learning resources may have an advantage over instructors: they can use their knowledge of their own previous misconceptions to create resources so there is less chance that they will suffer from an expert blind spot. However, it is very likely that some of the learning resources developed by students may be ineffective, inappropriate, or incorrect \cite{bates2014assessing}. As such, to effectively utilize resources developed by students, a selection and moderation process is needed to identify the quality of each resource. \name provides multiple options for moderation of the created resources. One of these options is ``staff moderation," where the created resources are moderated by the instructors before they are publicly released and added to the repository of the learning activities for the offering. However, this may not be feasible in large classes. Other options for moderation in \name rely on the collective wisdom of the crowd and methods that are generally used to review academic articles (e.g., the ``competent student moderation" option) or content moderation on social networks (e.g., ``flagging inappropriate content"). This raises an interesting research question: can non-experts accurately evaluate the quality of a learning resource? A paper by \citeA{Whitehill2019} 
provides evidence that non-expert subjective opinions may be able to accurately determine the quality of a learning resource, and that machine-learning algorithms may be used to infer the reliability of an individual's opinion, which further increases the accuracy of the results. Further investigations about this topic are underway by the authors. Our initial results are aligned with the findings of \citeA{Whitehill2019}.

Once a resource has been added to a course repository, students and instructors from that offering can view, attempt, and then evaluate the resources. For MCQs, once a user has attempted the question, they can view the right answer, the distribution of how the MCQ has been answered by their peers, and the explanation provided for the question. For all resources, users can identify the author and view the current rating and comments made about the resource. Users can add their own comments and rate the effectiveness of the resource.

\subsection{Learner Modelling} \label{sec:learnerModel}
In its current state, \name makes use of the Elo rating system, which was developed initially to rate chess players. In the educational context, the Elo rating system is employed to conduct a paired comparison among students and questions as competitors. If the question is answered correctly, the student's rating increases and the rating of the question decreases. If the question is answered incorrectly, the student's rating decreases and the rating of the question increases. The update to the ratings is proportional to the difference between the ratings of the student and the question. If a student who is highly rated correctly answers a low-rated question, then only a few rating points will be transferred from the low-rated question to the student. However, if a student with a low rating correctly answers a high-rated question, then many rating points will be transferred. One of the main benefits of using the Elo rating system in educational settings is its simplicity and self-correcting behaviour  \cite{pelanek2016applications}. In the standard Elo-based learner model, both students and items are considered as identical rivals and are both modelled by one parameter. In this model, the student parameter indicates the global proficiency level of the student on the entire domain. We make use of a multivariate Elo-based model \cite{Abdi2019Elo} that uses independent parameters to model a student's proficiency level on each individual concept in the domain. This is commonly referred to as using an ``overlaying" learner model, which is a widely accepted approach in modelling learners in \alss \cite{brusilovsky2007user, essa2016possible}.

Open learner models are learner models that are externalized and made accessible to students or other stakeholders, such as instructors, often through visualization, as an important means of supporting learning \cite{bull2010open}. They have been integrated into a variety of learning technologies, such as learning dashboards \cite{bodily2018open}, ITSs \cite{Ritter2007}, and \alss \cite{disalvo2014khan}, to help students and instructors monitor, reflect on, and regulate learning \cite{bull2010open}.

\begin {figure*}[h!]
\centering
\includegraphics[width=17.5 cm]{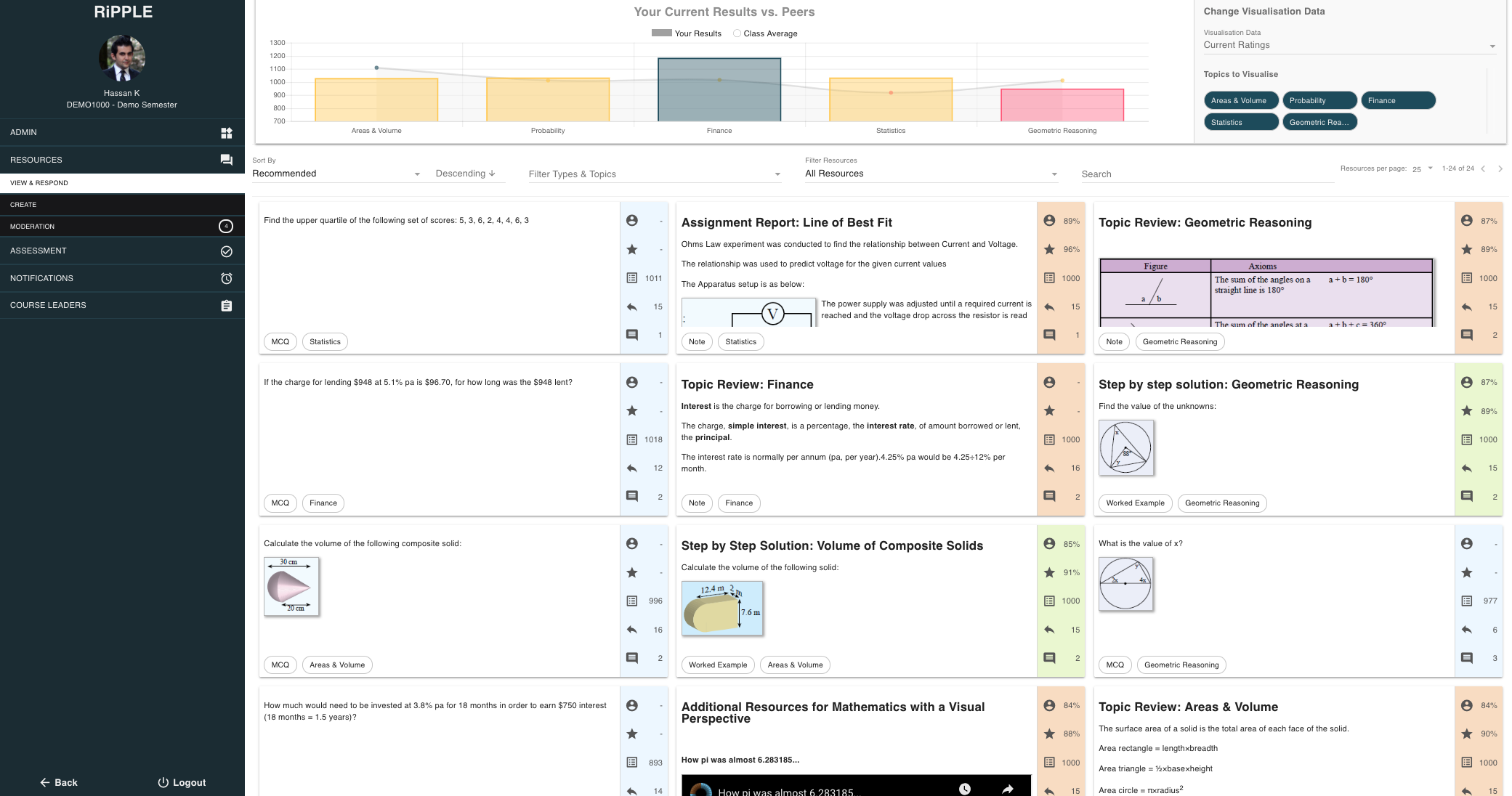}
\caption{Overview of open learner model and the resource selection and recommendation features in \name. \label{fig:overview}} 	
\end {figure*}

Figure~\ref{fig:overview} shows one of the main pages of \name, and in  particular its open learner model in the top panel. This interactive visualization widget enables students to view an abstract representation of their knowledge state based on the knowledge units present in the domain model. The proposed OLM was designed based on the following two principles: (1) The OLM and the recommendation results need to be placed close to one another on the interface. This was to help students better understand the rationale behind the recommendations of the system. (2) The utilized visualization must be easy to understand by the majority of the users. A range of visualizations --- including bar charts, line charts that demonstrate progress over time, zoomable Treemap, and Topic Dependency Models \cite{khosravi2018topic, cooper2018} --- have been incorporated and tested by earlier versions of \name. Based on the results of multiple rounds of usability tests, bar charts and line charts have been adopted in the latest version of the system. Accordingly, the ``Visualisation Data'' drop-down enables students to select between two models visualizing their knowledge state: viewing their current knowledge state and tracking changes to their knowledge state over time. The figure demonstrates how a student can view their current knowledge state. Each bar represents the competency of the student in one of the knowledge units of the course. The colour of the bars, determined by the underlying algorithm modelling the learner, categorizes competencies into three levels: red demonstrates inadequate competency in a topic, yellow demonstrates adequate competency with room for improvement, and blue demonstrates mastery in a knowledge unit. The model also shows the average competency of the entire cohort over each knowledge unit using a line graph.
 \citeA{Abdi2019Elo} show that opening the model in \name increases students' motivation to use the platform and increases their trust in the recommendations provided by the platform. RiPPLE's open learner model also provides additional insight for instructors on individual student-level or class-level gaps and competencies that can be used to improve item and course design.

\subsection{Resource Selection and Recommendation}\label{sec:recommendation}
The bottom part of Figure~\ref{fig:overview} presents the current graphical interface used for resource selection and recommendation in \name. A set of filters is available to help students search the resource repository. The  ``Filter Types \& Topics" option enables students to filter the resources included in the results based on their type (i.e., MCQs, worked examples, or notes) and tagged topics. This may help students target their learning toward particular topics or types of resources. The ``Filter Resources" option enables students to search for previously attempted (for revision), not attempted (for engaging with new content), and incorrectly answered (for reviewing content they previously found challenging) resources. This filter also enables students to view deleted resources they created themselves. This includes resources that they have personally deleted as well as those that have been deleted through moderation. Being able to access their deleted resources, students can revise and resubmit these resources. The ``Search" option enables students to search for resources based on specific content that may be present in the resource. The results of the search are presented as a list of learning resource cards that satisfy the specified filters.

The ``Sort By'' option allows students to sort the returned resources based on their difficulty, quality, number of responses, or personal fit (``Recommended"). If “Recommended” is selected, the platform sorts the resources based on their learning benefits to the student. This helps students make a good decision about what to study next \cite{Biggs1999}, while keeping them in the decision-making loop, thus maintaining autonomy. In its current state, \name employs the collaborative filtering algorithm proposed by  \citeA{khosravi2017} to create personalized recommendations for individual students that address their interests and their current knowledge gaps. The results reported in \citeA{khosravi2017} suggest that this algorithm outperforms baseline recommender systems in recommending learning resources.

Each resource card includes an overview of the resource content, the topics associated with the resource, and a sidebar in which the first icon shows the personal fit of the resource for the user (approximated by the recommendation engine). The second icon shows the general effectiveness of the resource (based on users' ratings of the resource). The third icon represents the difficulty level of the resource (approximated by the multivariate Elo-based model \cite{Abdi2019Elo}). The fourth icon shows the number of times the resource has been attempted, and the fifth icon presents the number of comments that have been written about the resource. Clicking on the resource card will take the user to another page that allows them to attempt and rate the resource.

\subsection{Student Profiles and Course Reports}\label{sec:studentProfiles}
It is increasingly being recognized that educational tools and technologies should not aim to replace instructors but rather provide support tools to enable instructors to improve their teaching practices \cite{collins2018rethinking}. As such, it is important for \alss and learning technologies more broadly to provide rich and timely actionable insights to instructors so that they can best manage their class within the context of their own course \cite{heffernan2014assistments}. It is equally important to provide rich feedback for students so that they can take ownership of their learning \cite{kitto2017designing}. In \name, student profiles and course reports are designed to provide rich analytics for instructors and students.

\paragraph{Student Profiles} 

Each student is provided with a personal profile that includes information on their achievements, engagement, and knowledge state. Instructors have access to the profiles of all of the students enrolled in their \name offering, which can help them identify the learning needs of each student. 
\name uses badges to help students track their progress. Students can achieve badges in two broad categories: ``Engagement Badges" and ``Competency Badges."

The engagement levels of students on a variety of tasks are presented using a visualization widget that enables them to use Kiviat diagrams, more informally known as radar charts, to compare their engagement against their peers. Kiviat diagrams have been used extensively in visualizing educational dashboards (e.g., see \citeA{may2011travis}) to display multivariate observations with an arbitrary number of variables \cite{chambers1983graphical}. Currently, engagement is computed based on the number of resources authored, resources answered, resources rated, and achievements earned.

\paragraph{Course Reports} \name allows instructors to download a set of course reports based on data collected by \name in their offering. These reports provide additional information about students, resources, comments, knowledge units, and students' attempts at resources. The data sets downloadable with these reports may be used to efficiently conduct educational research on many topics, including crowdsourcing, learner modelling, and recommender systems, with the consent of the users, at a low cost. Upon the first use of the platform, users are presented with a consent form that seeks their permission to use their data to improve our understanding of the learning process and to evaluate the effectiveness of the recommended content. \name enables users to change their response at any time. All users, regardless of their response, can use \name, and only data collected from users that have provided consent and have never changed their response will be used for research purposes. 


\section{Evaluation}\label{sec:evaluation}
An evaluation of the platform has been conducted using the following research questions:
\begin{itemize}[noitemsep]
    \item What measurable learning gains can be found when students use \name? 
    \item What evidence can be found that \name supports students in identifying resources that they think are effective?
    \item Do students perceive \name as beneficial to their learning?
\end{itemize}
To answer these questions, \name was piloted in an on-campus course on relational databases with 453 students at The University of Queensland\footnote{Approval from our Human Research Ethics Committee (\#2018000125) was received for conducting this evaluation on \name.}.
The course covers many topics that are generally included in an introductory course on relational databases, such as relational models, Structured Query Language (SQL), and database security. Data from the use of \name from the first five weeks of this course are used in this evaluation. To split students into the RiPPLE and non-RiPPLE groups, a threshold engagement of attempting at least 30 questions on \name was considered. Using this constraint, 215 (or 47\%) of the students were assigned to the RiPPLE group, and the remaining 238 students (or 53\%) were assigned to the non-RiPPLE group. Students in the RiPPLE group authored 351 MCQs (an average of 1.6 questions per student) and made 20,540 attempts (an average of 96 attempts per student).

\subsection{Measuring Learning Gains}\label{sec:measuring_gains}
In this section, we investigate whether using \name leads to measurable learning gains. We considered randomly assigning students to use \name versus a control, but this raised practical and ethical challenges \cite{sullivan2011getting}. In particular, since all students were in the same course, there was a high likelihood that some would realize there is a separate system, and it might conceivably be considered unfair to give one group the experience of \name over another. We therefore applied a quasi-experimental approach to analyzing students in the course who self-selected to engage (experimental group) with \name versus those who did not engage (control group). 

More precisely, we hypothesized that the choice of students to engage or not  with \name might be influenced by a set of their personal characteristics (covariates).  As a result, baseline covariates of students in the control group may be substantially different from those in the experimental group, which may bias the results of the investigation. We used propensity score matching (PSM) \cite{rosenbaum1983central}  to account for these potential differences in baseline covariates of the students. This method matches each student in the experimental group with a student from the control group such that the two students are similar based on their baseline covariates. For this experiment, current GPA, age, residency status (domestic or international), and program level (bachelor's level or master's level) of the students were selected as the covariates of the model. The PSM method using the potential outcomes framework, as described by \citeA{austin2011introduction}, was used to match students. A nearest-neighbour search was used to conduct a one-to-one matching and to find the smallest distance for the match within a predefined threshold ($\epsilon = 0.05$). Samples from the experimental group that did not have a match within the specified threshold in the control group were removed.

To cater for students' self-selection, the course used two rubrics to compute final grades. In one of the rubrics, the final exam and \name, respectively, had 40\% and 10\% contributions to the final grade. In the other rubric, the final exam and \name, respectively, had 50\% and 0\% contributions to the final grade. The final grade of a student was computed as the maximum of the values computed using the two assessment rubrics. The grade associated with \name included the following two criteria: 
\begin{itemize}[noitemsep]
    \item \textit{Answering and creating questions.} Students participated in a total of four rounds of authoring and answering questions. They received a total of two marks for each round of participation, once they correctly answered 10 questions (one mark) and created at least one correct and effective question (one mark). The effectiveness of questions was evaluated by the teaching staff and student peers.
\item \textit{Overall rating.} The overall rating of a student had a 2\% contribution to their grade. This was computed as an average rating calculated across all of the knowledge units set by the instructor. The mark associated with their overall rating was computed by $Min\left(Max\left(0, \frac{rating - 1000}{100}\right), 2\right)$.
The Min and Max functions are used to ensure a mark between 0 and 2.   
\end{itemize}

Grades from a midterm examination taken during week 7 were used to compare the learning gains of the control and experimental groups. The midterm contributed 15\% toward the final grade of the course. As also reported by \citeA{mojaradstudying}, comparisons of two possible breakdowns of \name students (experimental group) and non-\name students (control group) are presented:
\begin{enumerate}
    \item all \name students compared to all non-\name students, and
    \item \name students compared to matched non-\name students.
\end{enumerate}

 \begin {figure*}[h]
\centering
\includegraphics[width=17.4 cm, height=5.5 cm]
{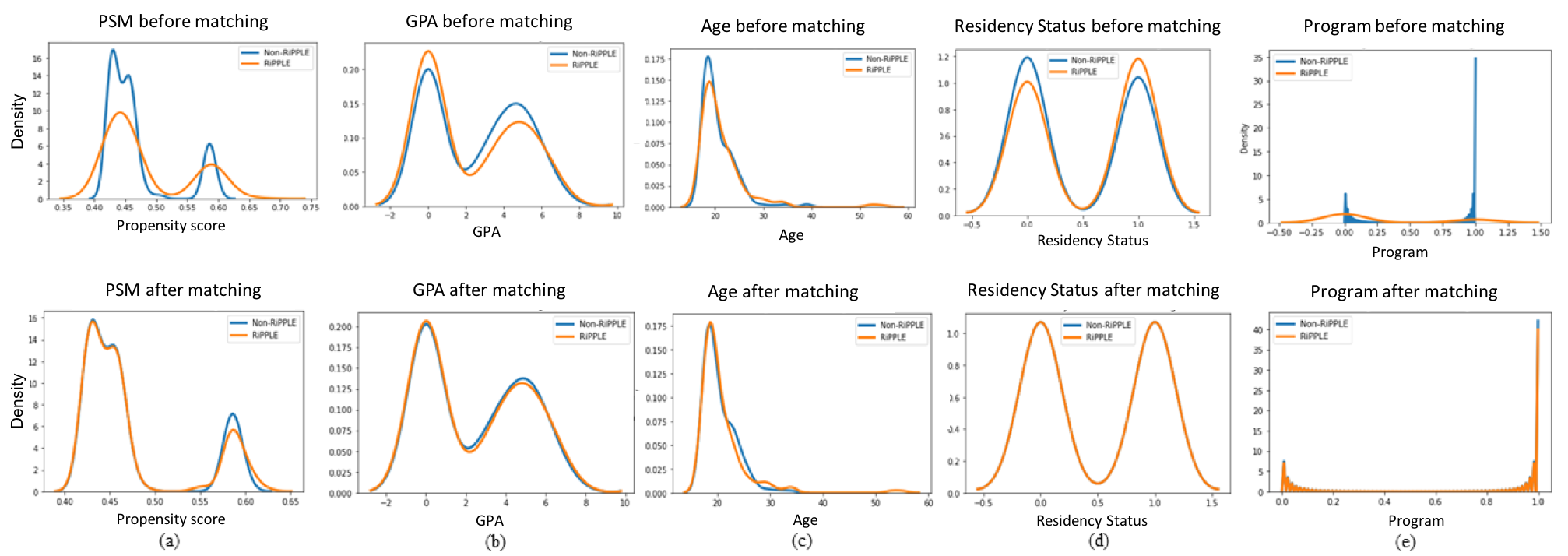}
\caption{Distribution of propensity score and covariates before and after matching. Table~\ref{tab:stats} reports the number of RiPPLE and non-RiPPLE members before and after matching. The figure illustrates that after matching, the two groups are similar across the four nominated covariates. \label{fig:comparePSM}} 	
\end {figure*}

Figure \ref{fig:comparePSM}(a)--(d) represents the distribution of each of the defined covariates (GPA, age, residency status, and program level) independently before and after matching, which clearly illustrates the effectiveness of PSM in diminishing the differences in the distribution of covariates between the experimental group and the control group. The overall propensity score of participants in both groups before and after matching is presented in Figure \ref{fig:comparePSM}(e),  which again illustrates that the participants in the experimental group were well matched to their control group counterparts.

Table \ref{tab:stats} summarizes the statistics related to midterm exam scores for each group before and after matching, where $\mu$ and $\sigma$ represent the average and standard deviation of the exam scores, respectively. A \textit{t}-test was used to evaluate the statistical significance of improvement in the learning gains. The average exam scores before matching for the students in the experimental group ($\mu = 73\% \pm 16\%$) were significantly higher than those of the students in the control group ($\mu = 65\% \pm 20\%$, $d=0.44$, $p < .001$). 
After matching, a total of 198 students were selected to participate in each group; 7 participants were removed from the experimental group because no match was found for them. Students' exam scores after matching revealed the same pattern, where the students in the experimental group had a statistically significantly better performance ($\mu = 74\% \pm 15\%$) than the students in the control group ($\mu=64\% \pm 21\%$, $d=0.54$, $p < .001$). The effect sizes of the two experiments support the claim that using \name does indeed provide measurable learning gains; however, the effect is considered to be at a medium level \cite{ma2014intelligent} and less effective than many of the studies discussed in Section~\ref{sec:relatedWork}.

\begin{table}[hbt]
\centering
\caption{Comparing the Exam Scores of \name and Non-\name Students}
\label{tab:stats}
\begin{tabular}{|c|c|c|c|c|c|c|c|}
\hline
\multirow{2}{*}{\textbf{Comparison}} & \multicolumn{3}{l|}{\textbf{\name}}                   & \multicolumn{3}{l|}{\textbf{Non-\name}}               & {\textbf{Effect size}}       \\ \cline{2-8}
                            & N & $\mu$ & $\sigma$                & N & $\mu$ & $\sigma$                & $d$                               \\ \hline
Non-matched                 & 215         & 73\%  &  16\% & 238         & 65\%  &  20\% & 0.44 \\ \hline
Matched                     & 198         & 74\%  &  15\% & 198         & 64\%  &  21\% & 0.54 \\ \hline
\end{tabular}
\end{table}

\subsection{Identifying Effective Resources}\label{sec:effectiveness}

\begin{figure}[h]
    \centering
    \includegraphics [width=6 cm] {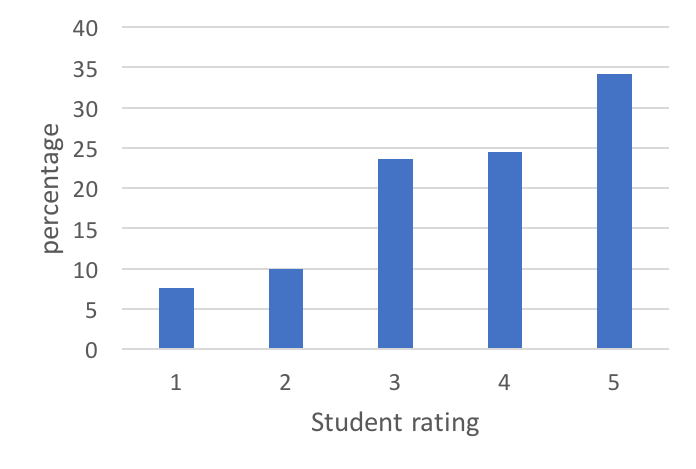}
   \caption{Students’ ratings of resources suggest that \name was successful in helping them identify resources that they think are effective.
 \label{fig:resourcesRatings}}
\end{figure}

This section investigates whether \name supports students in identifying resources that they think are effective. To do so, we employed the effectiveness ratings provided by the students themselves.  Figure~\ref{fig:resourcesRatings} shows that roughly 60\% of the time students provided a four- or five-star rating for a resource that they had attempted. Note in particular that five-star ratings were given 10 percentage points more than any other rating. In contrast, only around 17\% of ratings were  one or two star. This provides some evidence that the recommendation engine of \name was successful in enabling students to find resources that they thought were effective.

\subsection{Perceived Benefits for Students} \label{sec:perceivedBenefits}
A survey, advertised via an announcement on Blackboard, was conducted at the end of week 7 to capture students' perception of the platform. A total of 55 students completed the survey. The survey asked students to indicate their agreement with a set of statements using a five-point Likert scale. These statements were related to the following three broad categories: creation and evaluation of learning activities, visualization and recommendation, and the effectiveness of the platform itself. The actual statements and the corresponding label used here for each statement are presented in Table~\ref{tab:statement}, which is available in Appendix A.

\paragraph{Creation and Evaluation}

Figure~\ref{fig:survey-creation} shows survey results concerning student  perceptions about creation, completion, and evaluation of learning activities. Students reported that there were positive contributions to their learning from creating questions (63\% agreement vs. 13\% disagreement), writing solutions (63\% agreement vs. 13\% disagreement),  answering questions (67\% agreement vs. 12\% disagreement), and rating questions (56\% agreement vs. 16\% disagreement). In their written comments, a few students mentioned that a moderation process on the questions before they become available to the students would enhance the platform (``the moderation setting was set to no moderation for this course").
One student mentioned that a ``question originality/creativity" rating may help reduce the number of isomorphic questions.

\begin{figure}[h]
    \centering
    \includegraphics [width=12 cm] {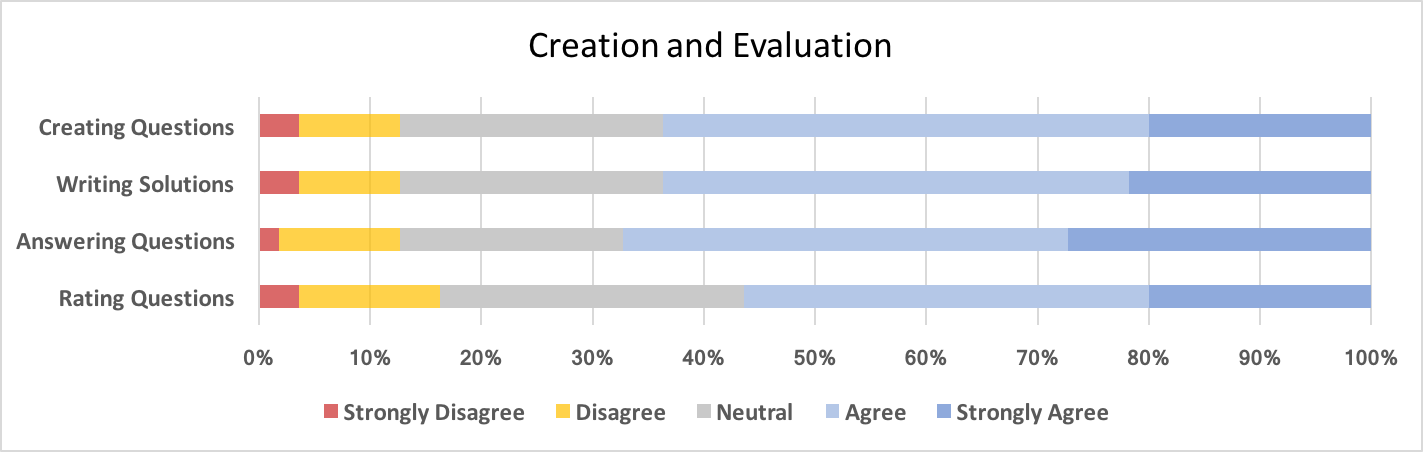}
   \caption{Results of the survey on questions related to creation and evaluation of learning activities. \label{fig:survey-creation}}
\end{figure}

\paragraph{Learner Model and Recommendation}

\begin{figure}[h]
    \centering
    \includegraphics [width=12 cm] {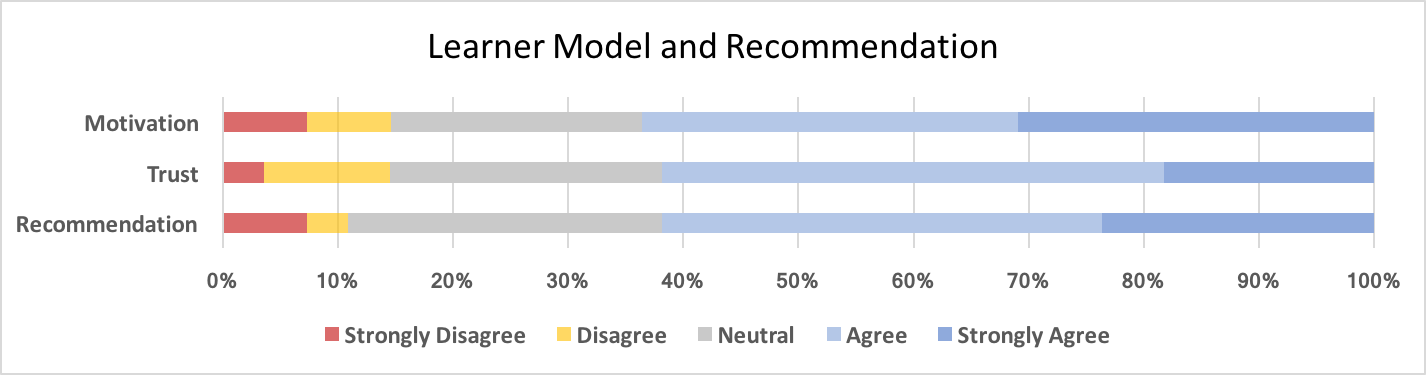}
   \caption{Results of the survey on questions related to the learner model and recommendation. \label{fig:survey-recommendation}}
\end{figure}

Figure~\ref{fig:survey-recommendation} shows survey results related to learner modelling and recommendation. Students reported that the open learner model increased their motivation to study or further use the platform (64\% agreement vs. 15\% disagreement) and increased their trust in the system (62\% agreement vs. 14\% disagreement). In their written comments, a few students mentioned that they would appreciate more transparency on how their knowledge state is computed, which is in line with recommendations from previous publications on open learner models \cite{Bull2016}. The majority of students indicated that the ``Recommend Question" feature helped them find effective questions (62\% agreement vs. 11\% disagreement). A few students identified a potential limitation that \name shares with many similar systems: because students cannot explicitly specify learning activities that they do not want to practise, they continued to have activities re-recommended despite purposely ignoring them in the past.

\paragraph{Platform Effectiveness}

\begin{figure}[h]
    \centering
    \includegraphics [width=12 cm] {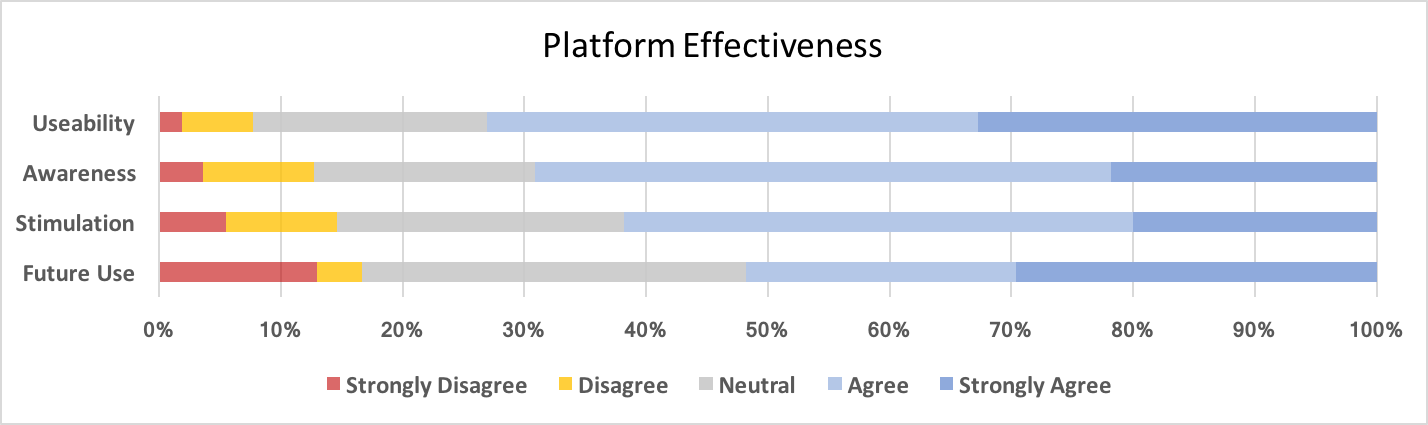}
   \caption{Results of the survey on questions related to the effectiveness of the platform. \label{fig:survey-effectiveness}}
\end{figure}

Figure~\ref{fig:survey-effectiveness} shows the  survey results relating to student-perceived effectiveness of the platform. Most students reported that they found the platform easy to use and navigate (73\% agreement vs. 8\% disagreement) and that it made them more aware of their current knowledge state (69\% agreement vs. 13\% disagreement). Most students also reported that \name stimulated them to study more effectively (62\% agreement vs. 15\% disagreement) and that they would like to use \name in other courses (52\% agreement vs. 17\% disagreement). Two limitations related to the general effectiveness of the platform were reported a number of times by students throughout the semester, which may have contributed to the lower level of interest in using the platform in other courses. The first limitation was related to the competency of students being decreased because of incorrectly answering a poorly worded or incorrect question. This limitation has now been resolved --- students will now regain their lost competencies from engaging with a poorly designed question once the question has been deleted from the platform. Another alternative for resolving this issue is to use the moderation feature. The second limitation relates to the lack of availability of an environment in \name where students can freely attempt questions without being concerned about their competencies being decreased. We are currently investigating ways that students' concerns about their competencies being decreased can be addressed. 

Most students perceived \name as beneficial to their learning; however, a significant number of students did not agree with the statements given in the survey. Focus groups and usability studies are underway to determine how we may further improve \name.

\section{Conclusion} \label{sec:conclusions}
This paper integrates insights from past research on crowdsourcing in education, learning sciences, and adaptive learning to present a platform called \name that can be used to collect data for LA research. Pedagogically, this tool encourages students to move from the passive acquisition of knowledge and skills to a more active learning mode, where they create learning activities for their peers. Thus, \name supports students in learning how to learn by working to improve their knowledge state through the recommendation of learning activities that best suit their learning needs, helping them visualize their progress along the way using open learner models. An initial evaluation of the platform in a class of 453 students has provided preliminary evidence that using \name (i) may lead to measurable learning gains and (ii) may support students in finding resources that they think are effective. Also, there is evidence that students perceived that using the platform benefited their learning.

There are, however, several limitations in the evaluation of the tool that restrict the generalizability of the results: (1) The quasi-experimental approach, even with the addition of propensity score matching, is prone to self-selection bias. Future studies aim to receive ethics approval to conduct a true randomized control trial experiment, seeking to provide more evidence in support of the claim that this tool results in demonstrable learning gains. (2) The current study explores the benefits of adopting \name using only one specific cohort: students with a background in computer science that have formal training in algorithmic literacy. Future work aims to investigate the benefits of adopting \name across more widely varying disciplines. This, in turn, may lead to the scaffolding of the platform or development of training material (e.g., short videos) so that students across all disciplines can use \name more effectively. (3) The current study only explores the benefits of using \name from the students' perspective. Future work aims to also explore the views of instructors that have used \name.

As a tool that is designed for easy integration (via LTI) with most standard LMSs, \name provides opportunities for both researchers and instructors to investigate the utility of applying crowdsourcing and adaptive learning methods in live teaching scenarios without requiring the ongoing and intensive generation of learning activities and sophisticated maps of how they interrelate. It is our hope that this will facilitate the wider use of an important technique for helping students actively participate in personalizing their learning experience while still acquiring skills and knowledge that are considered essential by instructors.  Thus, \name shows promise for advancing LA research in important areas, such as adaptive and personalized learning, student creativity and critical thinking, and learner modelling.

\subsection*{Declaration of Conflicting Interest}
The authors declared no potential conflicts of interest with respect to the research, authorship, and/or publication of this article.

\subsection*{Funding}
The authors declared no financial support for the research, authorship, and/or publication of this article.
\phantomsection

\addcontentsline{toc}{section}{Acknowledgments} 

\phantomsection
\bibliography{Master-references,kirstybib} 	
\newpage
\section{Appendix A}
A survey was conducted at the end of week 7 to capture students’ perception of the platform, with a total of 55 students responding. The survey asked students to use a five-point Likert scale to indicate their agreement with the statement given in Table~\ref{tab:statement}.

\begin{table}[h]
\caption{Statements Used in the Course Survey Described in Section~\ref{sec:perceivedBenefits}\label{tab:statement}}
\begin{tabular}{|l|p{13cm}|}
\hline
\multicolumn{2}{|c|}{\textbf{Creation and Evaluation}}                                                                                                                                       \\ \hline
Creating Questions  & Creating questions on \name helps me reflect on the learning objectives of the course.                                                         \\ \hline
Writing Solutions   & Writing solutions for my authored questions helps me develop a deeper understanding of the concept(s) targeted by the question.                                \\ \hline
Answering Questions & Answering questions on \name helps me develop a deeper understanding of the concept(s) targeted by the question.                               \\ \hline
Rating Questions    & Rating my peers' questions enables me to incorporate higher-order cognitive skills (e.g., appraise, judge, critique, support) in my own learning.              \\ \hline
Writing Comments    & Writing comments on my peers' questions enables me to incorporate higher-order cognitive skills (e.g., appraise, judge, critique, support) in my own learning. \\ \hline
\multicolumn{2}{|c|}{\textbf{Learner Model and Recommendation}}                                                                                                                              \\ \hline
Motivation          & The visualizations used by \name for showing my knowledge state increase my motivation to study or further use the platform.                   \\ \hline
Trust               & Having the ability to visually see my current knowledge state increases my trust in the recommended questions.                                               \\ \hline
Recommendation      & The recommend question feature helped me find effective questions.                                                                                            \\ \hline
\multicolumn{2}{|c|}{\textbf{Platform Effectiveness}}                                                                                                                                        \\ \hline
Usability          & \name is easy to use and navigate.                                                                                                             \\ \hline
Awareness           & \name makes me aware of my current learning situation.                                                                                                         \\ \hline
Stimulation         & \name stimulates me to study more effectively.                                                                                                 \\ \hline
Future use          & I would like to use \name in my other courses.                                                                                                 \\ \hline
\end{tabular}
\end{table}
\end{document}